\documentclass[aip]{revtex4-1}
\usepackage{graphicx}% Include figure files
\usepackage{dcolumn}% Align table columns on decimal point
\usepackage{bm}% bold math
\usepackage{amsmath}

\usepackage[utf8]{inputenc}
\usepackage[T1]{fontenc}
\usepackage{mathptmx}
\usepackage{etoolbox}
\usepackage{hyperref}
\hypersetup{
    colorlinks=true,
    linkcolor=blue,
    filecolor=magenta,      
    urlcolor=blue,
    pdfpagemode=FullScreen,
    citecolor=blue, 
    }
    
\makeatletter
\def\@email#1#2{%
 \endgroup
 \patchcmd{\titleblock@produce}
  {\frontmatter@RRAPformat}
  {\frontmatter@RRAPformat{\produce@RRAP{*#1\href{mailto:#2}{#2}}}\frontmatter@RRAPformat}
  {}{}
}%
\makeatother
\begin{document}

%\preprint{AIP/123-QED}

\title{Attaining high accuracy for charge-transfer excitations in non-covalent complexes at second-order perturbation cost: the importance of state-specific self-consistency}
% Force line breaks with \\
\author{Nhan Tri Tran}
\affiliation{University of Science, Vietnam National University, Ho Chi Minh City 70000, Vietnam}

\author{Lan Nguyen Tran}
\email{tnlan@hcmus.edu.vn}
\affiliation{University of Science, Vietnam National University, Ho Chi Minh City 70000, Vietnam}
\affiliation{International University, Vietnam National University, Ho Chi Minh City 70000, Vietnam}
\affiliation{Vietnam National University, Ho Chi Minh City 70000, Vietnam}

\date{\today}% It is always \today, today,

\begin{abstract}

Intermolecular charge-transfer (xCT) excited states important for various practical applications are challenging for many standard computational methods. It is highly desirable to have an affordable method that can treat xCT states accurately. In the present work, we extend our self-consistent perturbation methods, named one-body second-order M{\o}ller-Plesset (OBMP2) and its spin-opposite scaling variant, for excited states without additional costs to the ground state. We then assessed their performance for the prediction of xCT excitation energies. Thanks to self-consistency, our methods yield small errors relative to high-level coupled cluster methods and outperform other same scaling ($N^5$) methods like CC2 and ADC(2). In particular, the spin-opposite scaling variant (O2BMP2), whose scaling can be reduced to $N^4$, can even reach the accuracy of CC3 ($N^7$) with errors less than 0.1 eV. This method is thus highly promising for treating xCT states in large compounds vital for applications.

\end{abstract}

\maketitle

Various applications in modern technologies currently rely on intermolecular charge-transfer (xCT) excited states of non-covalent compounds, such as solar cells and optoelectronic devices \cite{vandewal2016interfacial, sarma2018exciplex, shen2021charge}. However, the xCT states are challenging for many widely-used methods. For example, time-dependent density functional theory (TD-DFT) and configuration interaction singles (CIS)\cite{HeadGordon:2005:tddft_cis}, which are perhaps the most affordable methods for excited-state calculations, are well known to underestimate considerably the excitation energies of xCT states\cite{HeadGordon:2005:tddft_cis}.
The major issue in these cases comes from the fact that significant changes of charge density require orbital relaxation that
is not described by the linear response of a Slater determinant.
\cite{subotnik2011cis_ct}
Even equation of motion coupled cluster theory with singles and doubles excitations (EOM-CCSD), \cite{Krylov:2008:eom_cc_review}
which is responding around a much more sophisticated wave function,
has difficulty in capturing these relaxation effects in many types of excitations
\cite{Krylov:2008:eom_cc_review,watts1996eomcc,Neuscamman:2016:var_qmc,Cheng2019eomKedge}. It is thus mandatory to optimize xCT state wavefunctions on an equal footing with ground-state ones. Recently, Kozma and coworkers \cite{kozma2020new} have systematically assessed the performance of coupled-cluster methods for xCT states' excitation energies. Those authors found that popular doubles methods like CC2, ADC(2), and EOM-CCSD are less accurate for xCT states than for valence states, and one needs to go beyond these methods to reach a satisfactory accuracy. However, the size of xCT compounds in practice is usually large, making high-level methods intractable. Thus, it is desirable to have an affordable method that can treat xCT states accurately.

To involve orbital relaxation effects in excited-state treatment, Gill and coworkers\cite{MOM-JPCA2008, MOM-JCTC2018} proposed the maximum overlap method (MOM) for self-consistent field (SCF), including HF and DFT. Instead of using the Aufbau principle, this algorithm maximizes the overlap between the occupied orbitals on successive SCF iterations. Several other algorithms optimizing excited states separately with ground state have been also developed, such as state-targeted energy projection (STEP)\cite{STEP-JCTC2020},  direct optimization\cite{levi2020variational}, and minimizing the square of the gradient\cite{hait2020excited, hait2021orbital, hardikar2020self}. In all those approaches, excitation energies are defined by the difference between excited-state and ground-state energies, resulting in the $\Delta-$SCF category. While $\Delta-$SCF with HF and DFT has been widely used and achieved a certain success in excited-state treatment \cite{macetti2021initial, daga2021electronic}, a natural question is how to incorporate more electron correlation into SCF solutions. Using a non-Aufbau determinant optimized via MOM-HF, Head-Gordon and coworkers have shown that the iterative coupled cluster can successfully describe some challenging excited states\cite{MOM-CC-2019, MOM-CC-2022}. However, the main concern for this approach is the convergence issues for excited states. While the non-iterative second-order M$\o$ller-Plesset perturbation theory (MP2) may provide an easier way, the accuracy of MP2 depends on the quality of the state-specific HF reference. We will show later that non-iterative MP2 correction fails to predict excitation energies of xCT states and is even worse than HF reference. Several previous works have applied MP2 to more sophisticated reference excited states involving many determinants, including the non-orthogonal configuration interaction MP2 (NOCI-MP2)\cite{noci_mp2_jcp2016,noci_mp2_jcp2018,noci_mp2_jcp2020} and excited-state MP2 (ESMP2)\cite{esmp2_jcp2018,esmp2_jctc2020,esmp2_jcp2023}. Despite their success, the use of sophisticated reference states complicates the subsequent perturbation treatment.

In the present work, we further extend our self-consistent perturbation methods, one-body MP2 (OBMP2) \cite{OBMP2-JCP2013,OBMP2-JPCA2021,OBMP2-PCCP2022,OBMP2-JPCA2023} and its spin-opposite scaling variant (O2BMP2)\cite{OBMP2-JPCA2024}, for excited-state treatment. To this end, we incorporate a state-targeting technique into our self-consistent perturbation methods to optimize excited-state wavefunctions. Currently, we employ the MOM approach to target the non-Aufbau determinant of the desired excited state during self-consistency. We apply our methods to evaluate the excitation energies of xCT states in various test sets. We have found that our self-consistent perturbation methods can outperform other same scaling ($N^5$) methods like MP2, CC2, and ADC(2). Noticeably, the spin-opposite scaling variant (O2BMP2), with a potential of scaling reduction to $N^4$, can reach the accuracy of triples coupled-cluster methods like CC3 with statistical errors less than 0.1 eV in many cases considered here. O2BMP2 is thus promising for treating xCT states in large compounds necessary for practical applications. 

%In the following, we will first briefly present the excited-state extension of our self-consistent perturbation methods using the maximum overlap method\cite{MOM-JPCA2008,MOM-JCTC2018}. We then test our implementation by evaluating the excitation energies of 150 valence transitions in a test set of singlet and doublet molecules. Subsequently, we move on to our primary goal, applying our methods to predict xCT excitation energies of various test sets reported by other authors and proposed by us.

The OBMP2 Hamiltonian is derived through the canonical transformation \cite{CT-JCP2006,CT-JCP2007,CT-ACP2007,CT-JCP2009,CT-JCP2010,CT-IRPC2010} followed by the cumulant approximation to reduce many-body operators into one-body operators\cite{cumulant-JCP1997,cumulant-PRA1998,cumulant-CPL1998,cumulant-JCP1999}. The OBMP2 Hamiltonian, whose derivation is presented in Refs.~\citenum{OBMP2-JPCA2021,OBMP2-PCCP2022,OBMP2-JPCA2023}, is given as
\begin{align}
  \hat{H}_\text{OBMP2} = \,\, \hat{H}_\text{HF} + \hat{v}_\text{OBMP2} \label{eq:h4}.
\end{align}
In Eq.~\ref{eq:h4}, $\hat{H}_\text{HF}$ is the standard HF Hamiltonian and $\hat{v}_\text{OBMP2}$ is a correlated potential operator composing of one-body operators and MP2 amplitude. The readers are referred to Refs.~\citenum{OBMP2-JPCA2021,OBMP2-PCCP2022,OBMP2-JPCA2023} for the working expression of the correlated potential operator.

We rewrite $\hat{H}_\text{OBMP2}$ in a similar form to standard HF as follows:
\begin{align}
  \hat{H}_\text{OBMP2} = & \hat{\bar{F}} + \bar{C}, \label{eq:h5}
\end{align}
where the constant $\bar{C}$ is a sum of terms without excitation operators. $\hat{\bar{F}}$ is the correlated Fock operator, $\hat{\bar{F}} =  \bar{f}^{p}_{q} \hat{a}_{p}^{q}$, with
correlated Fock matrix $\bar{f}^{p}_{q}$
\begin{align}
\bar{f}^{p}_{q} &= f^{p}_{q} + v^{p}_{q}. \label{eq:corr-fock}%, \\
\end{align}
$v^{p}_{q}$ is the matrix form of $\hat{v}_\text{OBMP2}$ and serves as the correlation potential altering the uncorrelated HF picture. The MO coefficients and energies then correspond to eigenvectors and eigenvalues of $\bar{f}^{p}_{q}$. We have recently introduced the spin-opposite scaling into the MP2 amplitude, resulting in the spin-opposite scaling variant (O2BMP2)\cite{OBMP2-JPCA2024}. Herein, we use the value of the spin-opposite scaling $c_{\text{OS}} = 1.2$ for all O2BMP2 calculations. 

To target a desired state during the OBMP2 self-consistency, we employ the MOM algorithm developed for HF and DFT \cite{MOM-JPCA2008, MOM-JCTC2018}. We first perform MOM-HF for desired states that are subsequently re-optimized using our methods. We also employed the DIIS technique to accelerate the convergence. Our methods, MOM-OBMP2 and MOM-O2BMP2, are implemented in a local version of PySCF\cite{pyscf-2018}. For comparison, we also carried out standard methods, including DFT, CIS, and coupled cluster, using PySCF and Orca package\cite{orca-2020}.

\begin{figure}[t!]
  \includegraphics[width=12cm]{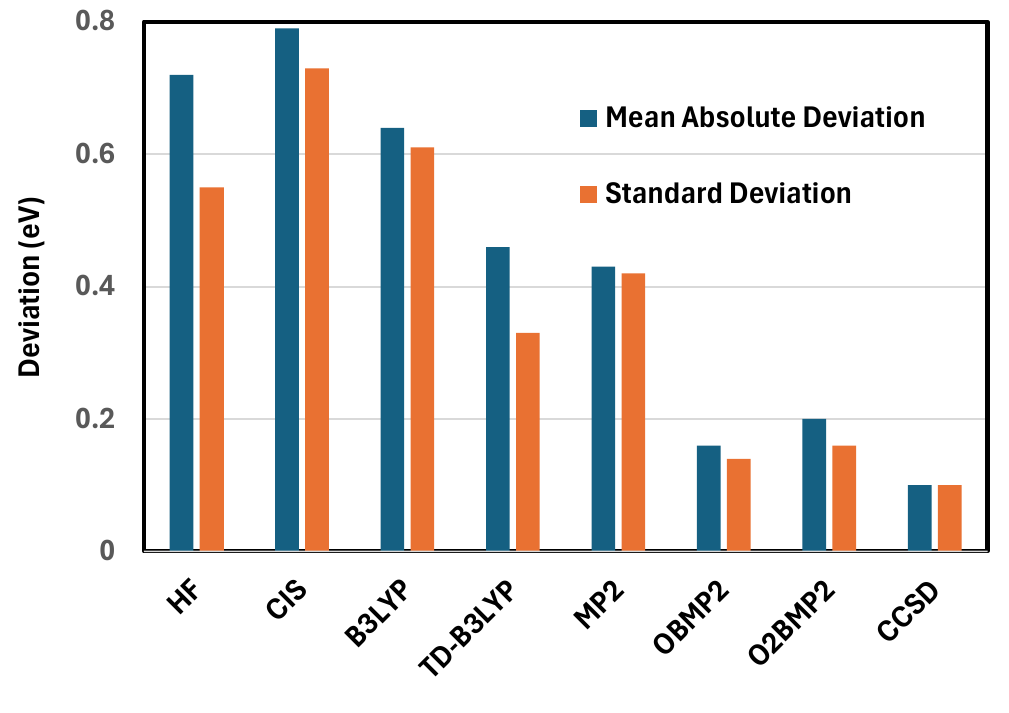}
  \caption{Maximum absolute deviation (MAD) and Standard deviation (SD) relative to CC3 for a test set of 152 valence transitions.}
  \label{fig:valence}
\end{figure}

We first test our implementation by considering 152 valence transitions of singlet and doublet molecules with geometries adapted from Refs.~\citenum{quest1} and ~\citenum{quest4}. We have employed the cc-pVDZ basis set \cite{DUNNING:1989:ccpvdz}. The raw data of all excitation energies are given in Supplementary Information (SI). In Figure~\ref{fig:valence}, we present mean absolute deviation (MAD) and standard deviation (SD) relative to the CC3 reference. EOM-CCSD slightly overestimates the CC3 reference with small MAD and SD (0.1 eV). We can see that both MOM-HF and MOM-B3LYP yield poor results with huge MADs and SDs. For valence states, state-specific SCF methods (MOM-HF and MOM-B3LYP) are inferior to their linear-response counterparts (CIS and TD-B3LYP). Although adding non-iterative MP2 correction on state-specific HF (MP2/MOM-HF) can lower errors, they are still quite large with MAD and SD up to $\sim$0.4 eV. As can be seen, OBMP2's statistical deviations ($\sim$0.15 eV) are much smaller than MP2's. Unsurprisingly, spin-opposite scaling (O2BMP2) is not superior to OBMP2 for valence states. Overall, our methods involving state-specific and second-order perturbative correlation effects simultaneously in self-consistency can improve upon non-iterative MP2 and reach the accuracy of coupled-cluster methods. This first assessment validates the excited-state extension of our methods. In the following, we move to the primary goal of this work, evaluating xCT excitation energies in various test sets of non-covalent compounds. 

\begin{table}[t!]
  \small
  \caption{\label{tab:CT14} \normalsize Relative Errors of Excitation Energies with Respect to CCSDT-3 for 13 xCT States given in Ref.~\citenum{kozma2020new}.}
  \begin{tabular}{ccccccccccccccc}
    \hline \hline		
Systems	                  &CT state\cite{kozma2020new}      &HF	    &MP2	&OBMP2	&O2BMP2	&EOM-CCSD\cite{kozma2020new}	\\
\hline
ammonia--fluorine	       &2$^1A_1$    &--1.75	&--1.11	&--0.50	&--0.57	&0.28\\
acetone--fluorine	       &3$^1A''$    &--2.16	 &0.24	&--0.17	&--0.22 &0.38\\
pyrazine--fluorine	       &2$^1B_2$    &--0.41	&--0.77	&--0.70	&--0.12 &0.44\\
	                        &2$^1A_2$     &--2.35	&0.43	&--0.05	&--0.29	&0.27\\
ammonia--oxygen difluoride. &4$^1A'$	   &--2.11	&0.01	&--0.52	&--0.39 &0.24\\
acetone--nitromethane	   &5$^1A$   &--1.98	&0.62  &0.28	&--0.02 &0.33\\
ammonia--pyrazine	       &5$^1A'$    &--0.91	&0.32	&--0.15	&--0.13 &0.37\\
pyrazine--pyrrole (H-bonded) &2$^1B_1$   &--1.13	&0.67   &0.08	&--0.18 &0.38\\
                            &2$^1A_1$   &--1.35	 &1.22	&0.09	 &--0.15 &0.32\\
	                         &3$^1A_1$  &--0.8	 &--1.08	&0.00	&--0.16 &0.30\\
pyrazine--pyrrole (stacked)	&2$^1A''$  &--1.11	&0.39	 &--0.26	&--0.28 &0.20\\
	                         &4$^1A'$ &--0.74	  &0.46	   &--0.10  &--0.14 &0.15\\
ethylene--tetrafluoroethylene &5$^1B_1$ &--0.22	&--1.1	&--0.38	&--0.51 &0.29\\
\hline
MAD		&&1.40	&0.61	&0.24	&0.24	&0.30\\
SD      &&0.70	&0.38	&0.21	&0.16	&0.08\\
\hline \hline
\end{tabular}
\end{table}

\begin{figure}[t!]
  \includegraphics[width=10cm]{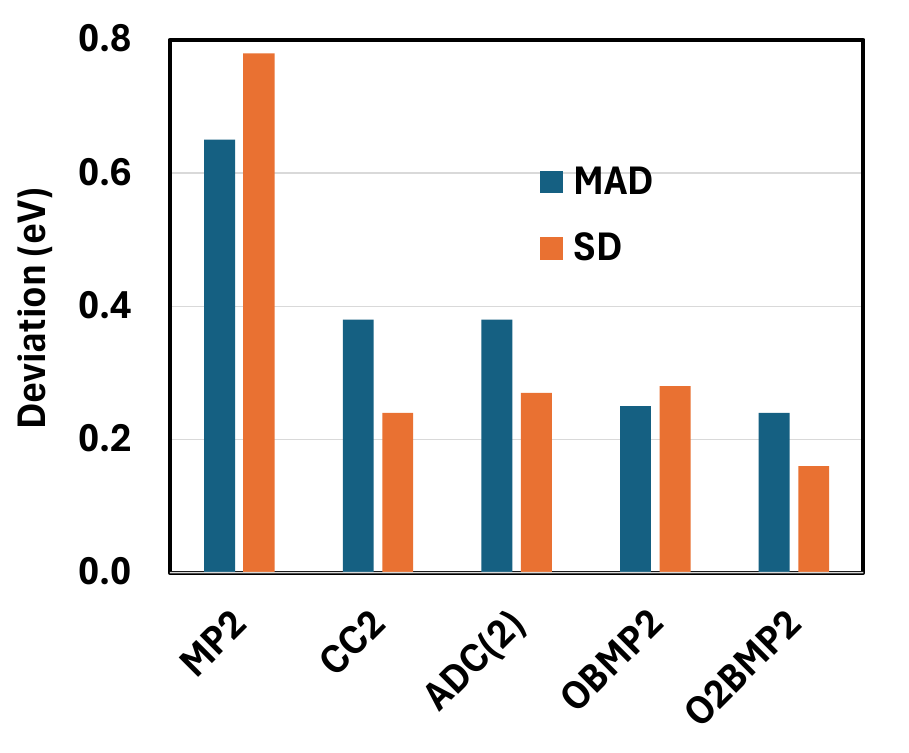}
  \caption{Maximum absolute deviation (MAD) and Standard deviation (SD) relative to CCSDT-3 of different $N^5$ methods for a test set of 13 xCT states adopted from Ref~\citenum{kozma2020new}. CC2 and ADC(2) results are taken from Ref~\citenum{kozma2020new}.}
  \label{fig:xCT13}
\end{figure}

Kozma and coworkers have performed systematically a benchmark for coupled-cluster methods on a test set of xCT states\cite{kozma2020new}. We thus adapt this test set to compare our methods with other same scaling methods as well as high-level coupled-cluster methods. The raw data of excitation energies are presented in SI, and relative errors to CCSDT-3 are reported in Table~\ref{tab:CT14}. We can see that MOM-HF cannot predict xCT excitation energies accurately. It significantly underestimates xCT excitation energies with huge MAD (1.40 eV), implying that the orbital optimization with only state-specific effects is insufficient to reach high accuracy for xCT states. Adding dynamical correlation on state-specific HF (MP2/MOM-HF) can reduce MAD and SD more than twice, but they are still enormous. Not surprisingly, EOM-CCSD, adapted from Ref.~\citenum{kozma2020new}, with higher scaling yields errors smaller than MP2. When the orbital optimization considers both state-specific and correlation effects simultaneously, errors are dramatically improved upon MP2, and our methods' MADs (0.24 eV) are even smaller than the EOM-CCSD one (0.30 eV). 

To see the superiority of our methods, we compare them with other methods having the same scaling ($N^5$) reported in Ref.~\citenum{kozma2020new} and present MADs and SDs relative to CCSDT-3 in Figure~\ref{fig:xCT13}. As can be seen, CC2 and ADC(2) with more sophisticated formulations than non-iterative MP2 correction on state-specific HF wavefunctions can yield much smaller errors. However, when the second-order perturbation correction is self-consistently updated in our method, it can outperform linear-response methods like CC2 and ADC(2). We would like to emphasize that the excited-state extension of our methods is simple and has the same scaling as the ground state. While the spin-opposite scaling variant (O2BMP2) yields a similar MAD with OBMP2, it improves SD upon OBMP2. In general, for this test set, our methods can outperform same-level methods and reach the accuracy of EOM-CCSD. 

\begin{table}[t!]
  \small
  \caption{\label{tab:Eric_CT} \normalsize Relative Errors of Excitation Energies with respect to $\delta$-CR-EOM-CC(2,3),A for a test set of 15 xCT States adapted from Ref.~\citenum{tuckman2024aufbau}.}
  \begin{tabular}{ccccccccccccccc}
    \hline \hline		
Systems	                    &state$^a$ 	&HF	    &MP2	&OBMP2	 &O2BMP2\\
\hline
ammonia--difluorine 	    &h1p1	&--1.41	&--3.04	&--0.07	 &--0.05	\\
chloride--carbon monoxide	&h1p2	&--0.69	&--0.72	  &0.21	 &--0.06	\\
                        	&h1p1	&--0.69	 &0.21	  &0.19	 &--0.04	\\
                        	&h2p2	&--0.70	&--0.71	  &0.18	 &--0.06	\\
                        	&h2p1	&--0.70	&--0.73	  &0.18	 &--0.05	\\
                        	&h3p2	&--0.70	&--0.71	  &0.17	 &--0.06	\\
                        	&h3p1	&--0.71	  &0.20	  &0.18	 &--0.06	\\
chloride--dinitrogen 	    &h1p2	&--0.69	&--0.67	  &0.17	 &--0.06	\\
                        	&h1p1	&--0.84	  &0.30	  &0.21	 &--0.02	\\
                        	&h3p1	&--0.83	  &0.28	  &0.20	 &--0.04	\\
                        	&h2p1	&--0.82	  &0.29	  &0.18	 &--0.03	\\
                        	&h3p2	&--0.82	  &0.29	  &0.18	 &--0.03	\\
                        	&h2p2	&--0.83	  &0.29	  &0.20	 &--0.03	\\
chloride--ethylene      	&h1p1	&--0.53	&--0.01	&--0.07	 &--0.09	\\
                        	&h3p1	&--0.46	  &0.03	  &0.01	 &--0.05	\\
\hline
MAD		&&0.76	&0.57	&0.16	&0.05	 \\	
SD		&&0.21	&0.73	&0.06	&0.02	\\
\hline \hline 
\\
\end{tabular}\\
$^a$\footnotesize{See Ref.~\citenum{tuckman2024aufbau} for the notation of states.}
\end{table}

\begin{figure}[t!]
  \includegraphics[width=10cm]{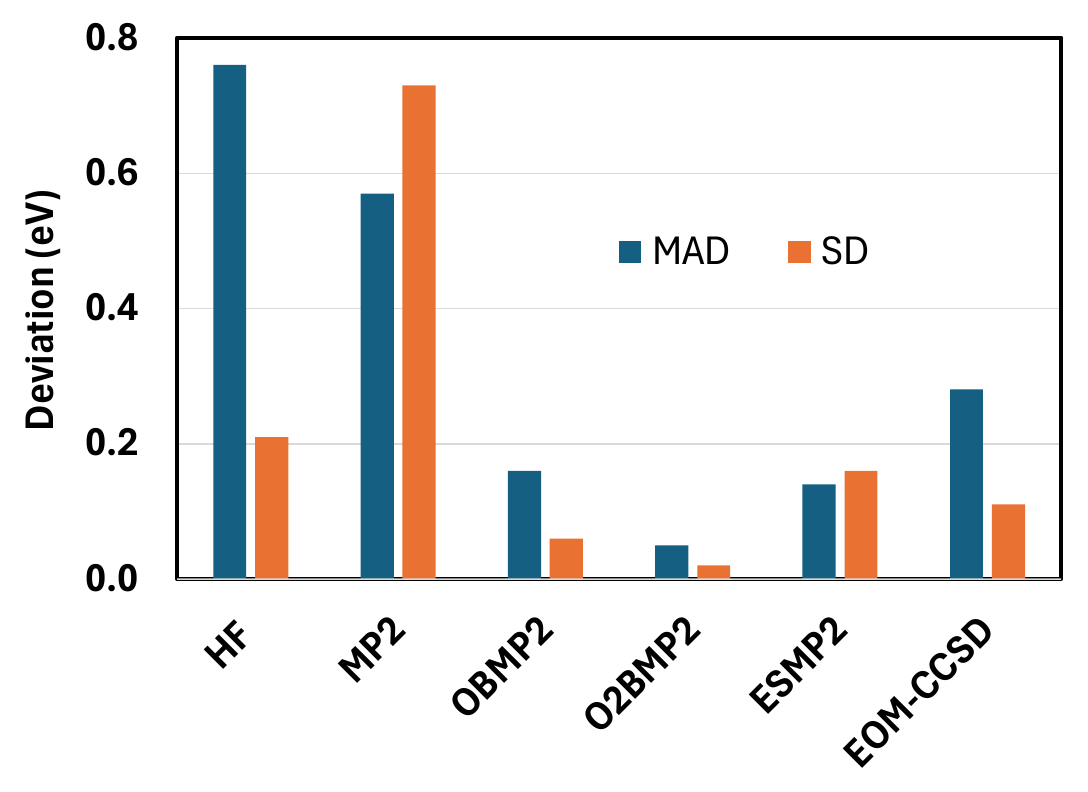}
  \caption{Maximum absolute deviation (MAD) and Standard deviation (SD) relative to $\delta$-CR-EOM-CC(2,3),A of different methods for a test set of 16 xCT states adopted from Ref~\citenum{tuckman2024aufbau}. ESMP2 and EOM-CCSD results were also taken from Ref~\citenum{tuckman2024aufbau}.}
  \label{fig:cteric}
\end{figure}

Recently, Tuckman and Neuscamman have shown that excited-state second-order perturbation theory (ESMP2) with a scaling of $N^5$ can outperform EOM-CCSD with a higher scaling, $N^6$, for a test set of 15 xCT states\cite{tuckman2024aufbau}. ESMP2 is a non-iterative second-order perturbation correction to excited-state mean-field (ESMF) reference similar to MP2 for ground state\cite{esmp2_jcp2018,esmf_jctc2020}. However, the formulation of ESMP2 is much more complicated than MP2 due to the sophistication of ESMF wavefunction. Thus, it is exciting to use the test set reported in Ref~\citenum{tuckman2024aufbau} to assess the performance of our self-consistent second-order perturbation theory implemented without any complicated modification from the ground state. Table~\ref{tab:Eric_CT} presents relative errors with respect to $\delta$-CR-EOM-CC(2,3),A\cite{tuckman2024aufbau}. The raw data of excitation energies from HF, MP2, and our methods are given in SI. For comparison, we also adapted ESMP2 and EOM-CCSD results reported in Ref.~\citenum{tuckman2024aufbau} and plotted them together with HF, MP2, and our methods in Figure~\ref{fig:cteric}. In the state notation,``h" and ``p" stand for hole and particle, respectively. The readers see Ref.~\citenum{tuckman2024aufbau} for particle and hole orbitals defined xCT states. 

We can see that MP2/MOM-HF does not improve MOM-HF results, and MP2's SD is much larger than HF's. The failure of MP2 again indicates that non-iterative correction of dynamical correlation on state-specific HF wavefunctions is insufficient to describe xCT states. EOM-CCSD, with the lack of full orbital relaxation, as argued in Ref.~\citenum{tuckman2024aufbau}, provides significant errors with MAD up to 0.28 eV. ESMP2 with a more sophisticated reference can achieve MAD two times smaller than EOM-CCSD. Interestingly, OBMP2 involving state-specific and dynamical correlation effects during the orbital optimization also yields MAD and SD smaller than EOM-CCSD. Although OBMP2's MAD is similar to ESMP2's, its SD is smaller than the latter. Excitingly, spin-opposite scaling (O2BMP2) can reduce errors further to 0.05 eV and 0.02 eV for MAD and SD, respectively. Overall, our excited-state method, without any significant modification from the ground-state one, can yield errors similar to a more sophisticated method, ESMP2, and even better than the latter when spin-opposite scaling is employed. 

\begin{table}[t!]
  \small
  \caption{\label{tab:x40} \normalsize Relative Errors of Excitation Energies with Respect to CC3 for 19 xCT States in 11 non-covalent systems.}
  \begin{tabular}{ccccccccccccccc}
    \hline \hline		
Systems	                &State	&HF	    &MP2	&OBMP2	&O2BMP2	&EOM-CCSD	\\
\hline

methane--F$_2$	            &h1p1	&--1.44	&--2.35	&--0.34	&--0.13	&0.22\\
	                    &h2p1	&--1.48	&--2.39	&--0.30	&--0.17	&0.22\\
methane--Cl$_2$	        &h1p1	&  0.47 &  0.18	&--0.27	&  0.00	&0.15\\
	                    &h1p2	&  0.30	&  0.03	&--0.24	&  0.17	&0.09\\
fluoromethane--methane	&h1p1	&--0.46	&--0.69	&--0.11	&--0.10	&0.04\\
chloromethane--methane	&h1p1	&--0.54	&--0.75	&--0.25	&--0.20	&0.04\\
HF--methanol	        &h1p1	&--0.73	&--0.59	&  0.16	&  0.00	&0.07\\
HCl--methanol	        &h1p1	&--0.19	&--0.48	&--0.07	&--0.01	&0.16\\
	                    &h1p2	&--0.21	&--0.44	&--0.01	&  0.04	&0.13\\
                    	&h1p3	&--0.39	&--0.59	&  0.02	&--0.05	&0.16\\
	                    &h2p3	&--0.38	&--0.58	&  0.02	&--0.04	&0.16\\
HF--methyl amine	    &h1p1	&--0.73	&--0.71	&  0.05	&--0.12	&0.06\\
HCl--methyl amine	    &h1p1	&--0.13	&--0.47	&--0.10	&--0.03	&0.20\\
water--ammonia	        &h1p1	&--0.76	&--0.88	&  0.14	&--0.09	&0.18\\
                    	&h1p2	&--0.61	&--0.72	&  0.08	&--0.04	&0.11\\
                    	&h1p3	&--0.75	&--0.87	&  0.14	&--0.08	&0.16\\
water dimer         	&h1p1	&--0.70	&--0.82	&  0.06	&--0.08	&0.07\\
water--methane      	&h1p1	&--0.63	&--0.89	&--0.04	&--0.1	&0.07\\
                    	&h2p2	&--0.70	&--0.70	&  0.18	&--0.02	&0.02\\
 \hline                     
MAD		                       &&0.61	&0.80	&0.14	&0.08	&0.12	\\
SD		                       &&0.36	&0.60	&0.10	&0.06	&0.06	\\
\hline \hline
\end{tabular}
\end{table}

\begin{figure}[t!]
  \includegraphics[width=12cm]{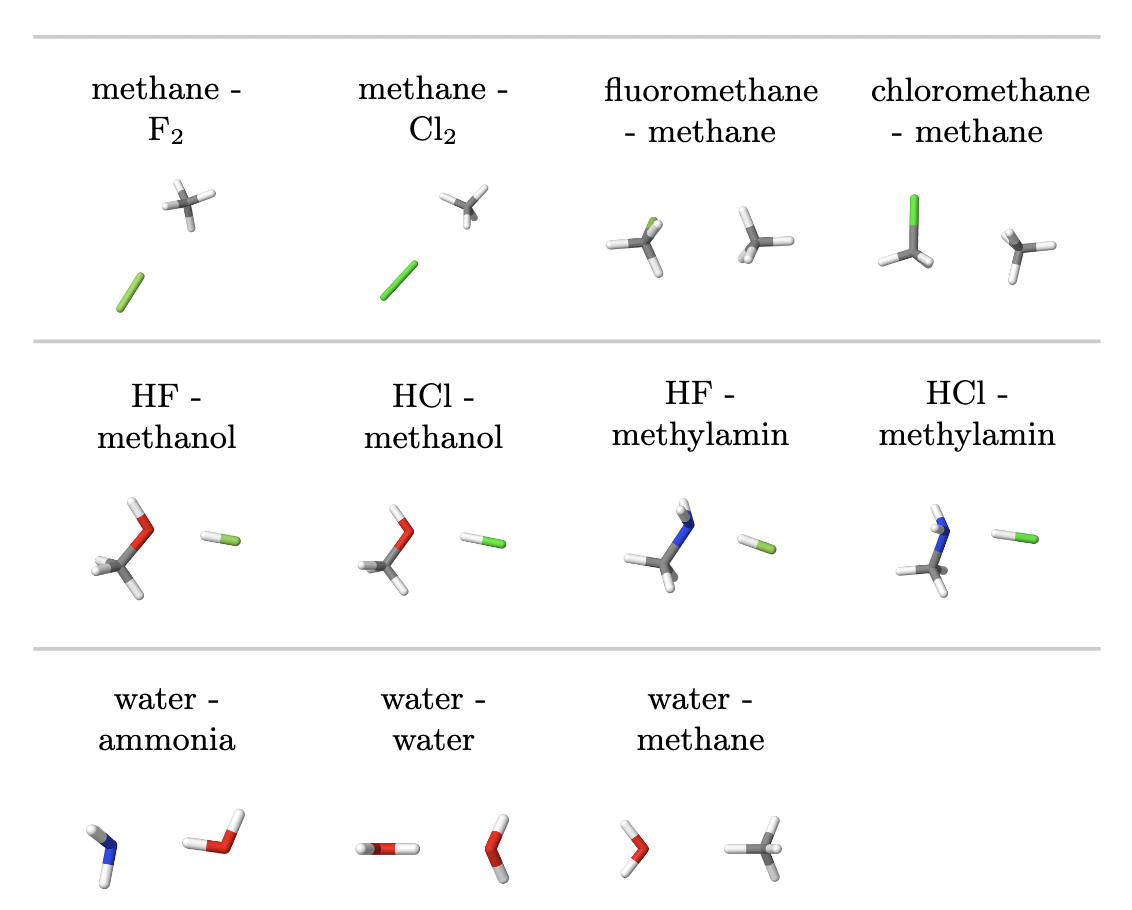}
  \caption{11 non-covalent systems used in Table~\ref{tab:x40}.}
  \label{fig:new_xct}
\end{figure}

We now assess the performance of our methods on a new set consisting of 19 xCT states of 11 non-covalent systems depicted in Figure~\ref{fig:new_xct}. Molecular geometries were taken from Refs.~\citenum{X40-testset,A20-testset} and the cc-pVDZ basis set was used. Orbitals defined states and all excitation energies are given in SI. We employ CC3 as the reference for this test, and errors relative to it are presented in Table~\ref{tab:x40}. Overall, while EOM-CCSD overestimates excitation energies, other methods underestimate them. Again, we can see that non-iterative MP2 correction does not help improve state-specific HF results and is even worse than the latter. EOM-CCSD with higher scaling can yield much smaller errors with MAD and SD of 0.12 eV and 0.06 eV, respectively. Interestingly, OBMP2 can reach EOM-CCSD's accuracy and when the spin-opposite scaling (O2BMP2) is exploited, the errors are further reduced and even smaller than EOM-CCSD ones. Looking closely, we see that the vast errors of MP2 are mainly due to the two states of methane--F$_2$. The poor description of state-specific HF wavefunctions may deteriorate the MP2 performance. On the contrary, the state-specific self-consistency in our methods makes them adequately describe the xCT state wavefunctions, leading to small errors with respect to higher-level methods.

\begin{figure}[t!]
  \includegraphics[width=10cm]{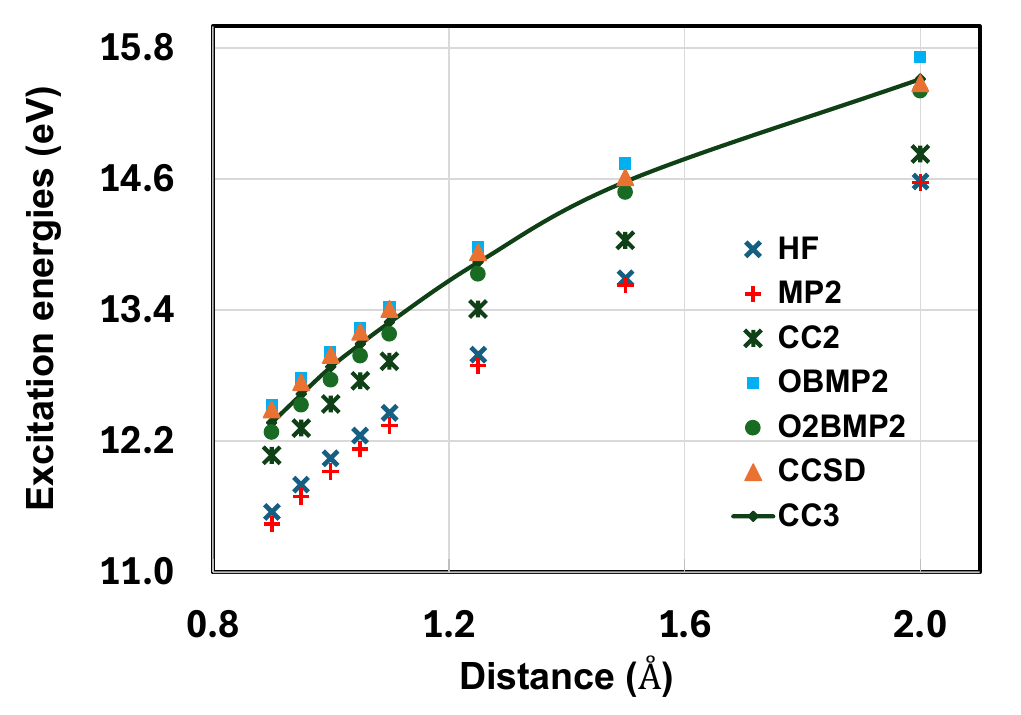}
  \caption{Excitation energies of xCT states in water dimer as a function of water--water distance.}
  \label{fig:waterdimer}
\end{figure}

Let us assess whether our methods are consistently good when the distance between two monomers increases. In Figure~\ref{fig:waterdimer}, we plot the xCT excitation energy of the water dimer as a function of water--water distance. We use CC3 as the reference and show CC2 for comparison. It is not surprising that as the distance increases, the excitation energies increase. As can be seen, HF and MP2 notoriously underestimate excitation energies at different distances with huge errors. The poor description of state-specific HF wavefunction may cause the failure of non-iterative MP2 correction. The CC2 method, with the same scaling as MP2 but a more sophisticated formulation, can provide better performance at short distances. However, CC2 deviates far from the CC3 reference at long distances. EOM-CCSD, with higher scaling than CC2, yields better results and is very close to CC3. Our methods, OBMP2 and O2BMP2, are in good agreement with CC3 for the whole range of distances. However, as the distance becomes longer, while the OBMP2 deviation from CC3 slightly increases, exploiting the spin-opposite scaling makes the deviation smaller and excitation energies consistently close to CC3 references. In general, O2BMP2 with a potential of scaling reduction to about $N^4$ can reach the accuracy of higher-scaling methods in predicting xCT excitation energies even at long distances.

In summary, we have extended our self-consistent perturbation methods, OBMP2 and O2BMP2, to excited-state treatment without additional costs to the ground-state one. We then assessed their performance in predicting intermolecular CT excitations, which are challenging for many linear-response approaches. While non-iterative MP2 correction to state-specific HF reference cannot improve HF results, our self-consistent methods can dramatically lower the errors for all cases considered here. Interestingly, our self-consistent perturbation methods can outperform other same scaling ($N^5$) methods like CC2 and ADC(2). In particular, the spin-opposite scaling variant (O2BMP2), with a potential of scaling reduction to $N^4$, can reach the accuracy of high-level coupled cluster methods, such as EOM-CCSD ($N^6$) and CC3 ($N^7$). We thus expect that O2BMP2 is highly promising for treating xCT states in large compounds necessary for practical applications.

\section*{Acknowledgment}
This research is funded by Vietnam National University HoChiMinh City (VNU-HCM) under grant number C2024-28-04.

\section*{Supplementary Information}
Raw data for all excitation energies (in eV) calculated in this work and orbitals defined 19 xCT states in Table~\ref{tab:x40}.

\bibliography{main}
\end{document}